# Electron emission properties of two-phase argon and argon-nitrogen avalanche detectors

A. Bondar, A. Buzulutskov[*], A. Grebenuk, D. Pavlyuchenko, Y. Tikhonov

*Budker Institute of Nuclear Physics, Lavrentiev avenue 11, 630090 Novosibirsk, Russia*
*E-mail:* `A.F.Buzulutskov@inp.nsk.su`

ABSTRACT: Electron emission properties of two-phase Ar avalanche detectors are studied. The detectors investigated comprised a liquid Ar or Ar+$N_2$ layer followed by a multi-GEM multiplier operated in the saturated vapour at 84 K. Two components of the electron emission through the liquid-gas interface were observed: fast and slow. In Ar, the slow emission component dominated even at higher fields, reaching 2 kV/cm. In Ar+$N_2$ on the contrary, the fast emission component dominated at higher fields, the slow component being disappeared. This is explained by the electron backscattering effect in the gas phase. The slow component decay time constant was inversely proportional to the electric field, which is compatible with thermionic emission model. The electron emission efficiencies in two-phase Ar and Ar+$N_2$ were estimated to be close to each other.

---

[*] Corresponding author.

# Contents

## 1. Introduction

Two-phase emission detectors, with electron emission through the liquid-gas interface, have been known since the seventies [1],[2]. Two-phase emission detectors with optical readout, using PMTs and proportional scintillations in the gas phase, have been already applied in dark matter search [3],[4],[5] and coherent neutrino-nucleus scattering experiments [6],[7].

Not long ago two-phase avalanche detectors, operated in electron avalanching mode, have been introduced [8],[9]. In such detectors the signal is recorded using hole gas multipliers, namely Gas Electron Multipliers (GEMs) [10] or thick GEMs (THGEMs) [11],[12], operated in saturated vapour above the liquid phase. Most promising results were obtained with two-phase Ar avalanche detectors providing gains reaching $10^4$, using GEMs [13],[14],[15] or THGEMs [16],[17],[18], and detecting both primary scintillation and ionization signals using CsI photocathodes [14],[15]. At present two-phase Ar avalanche detectors are started using in coherent neutrino-nucleus scattering [19] and dark matter search [20] experiments. They might also have applications in large-scale detectors for long-baseline and cosmic neutrinos [21]. Two-phase Xe avalanche detectors exhibiting lower gains, of the order of 100 [13],[22], might be attractive for Positron Emission Tomography (PET) [9]. Two-phase He and Ne avalanche detectors were proposed for solar neutrino detection [23].

One of the most interesting features of two-phase systems is the effect of electron emission from the liquid to the gas phase [24],[25],[26]. In two-phase Ar such process has fast and slow electron emission components, lasting for less than a nanosecond and larger than few microseconds respectively [16],[24],[25]. The fast component was explained by emission of "hot" electrons heated by an electric field during drifting in the liquid and having overcome a potential barrier at the liquid-gas interface [27],[28]. The slow component was explained by thermionic emission of "cold" electrons including those cooled down after a reflection from the potential barrier [25],[27]. It should be emphasized that the slow component was never observed in two-phase Kr and Xe [27],[28], presumably due to a higher potential barrier as compared to Ar. And vice versa, the fast component was never observed in two-phase He and Ne [26],[27], since there the electrons are localized in bubbles and thus cannot be heated by the electric field. Accordingly, the two-phase Ar system is the unique one providing an opportunity to study both fast and slow electron emission processes, the electron emission effect being taken as a whole.

In previous works the electron emission properties of two-phase Ar were studied separately: either the emission efficiency was measured but the time characteristics were missed [24] or the slow component time constant was measured but the emission efficiency characteristics were missed [25]. This was because in those works the internal amplification was not possible and the rather large amplifier shaping time (400 μs) was used, preventing from direct observation of both fast and slow components. Accordingly the direct observation of the fast and slow emission components has become possible only when the avalanching of emitted electrons in two-phase detectors had been realized with the help of the fast GEM multipliers [16].

In the present work we study in detail electron emission properties of two-phase Ar avalanche detectors based on GEMs using a pulse shape and spectrum analysis. These properties include the fast and slow component fractions, slow component time constant and electron emission efficiency, measured as a function of the electric field. Apparently the long signal due to the slow component may not be desirable in time measurements and high flux environment. In connection with this, the effect of doping Ar with $N_2$ on electron emission properties in two-phase avalanche detectors, and in particular a possibility to suppress the slow component, has for the first time been studied (scintillation and ionization detection properties in liquid Ar doped with $N_2$ were studied elsewhere [29],[30],[31]). The physics of electron emission in two-phase systems is also considered.

## 2. Experimental setup

The experimental setup and procedures were described elsewhere [13],[14],[15],[16]. Here we describe details relevant to the measurements of the electron emission properties. The cryogenic chamber had a volume of 2.5 l and comprised a cathode mesh at the chamber bottom immersed in the liquid and a multi-GEM assembly of an active area of 3×3 cm$^2$ placed in saturated vapour above the liquid. The first electrode of the first GEM was in addition coated with a CsI photocathode. The distance between the cathode and the first GEM was 11 mm and between the GEMs - 2 mm.

In the two-phase mode the detector was operated in Ar in equilibrium state at a point a little above the triple point, namely at a saturated vapour pressure of 0.70 atm corresponding to a temperature of 84 K [32]. The total liquid layer thickness was 10 mm. In the cathode gap the liquid and gas layer thicknesses were 8 and 3 mm respectively.

The cathode and GEM electrodes were biased through a resistive high-voltage divider placed outside the cryostat. The multi-GEM multiplier was operated in either a triple-GEM or double-GEM readout mode, namely in either 3GEM or 2 GEM mode. Accordingly, the anode signals were read out from the last electrode of either the third or the second GEM respectively, using a charge-sensitive preamplifier with a 10 ns rise time and sensitivity of 0.5 V/pC followed by a research amplifier. The amplification factor of the latter was 20 and its shaping time was either 0.5 or 10 μs. The signals were analyzed with a TDS5032B digital oscilloscope.

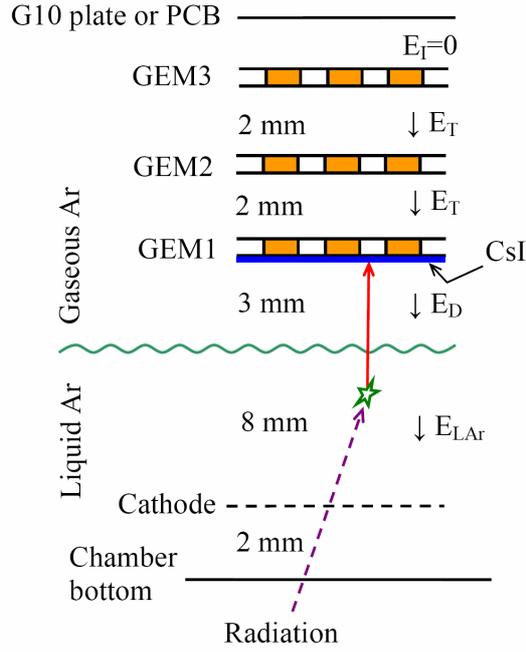

Fig. 1. Schematics view of the experimental setup to study electron emission properties of two-phase Ar avalanche detectors.

In the present work the emission properties of two-phase detectors were studied as a function of the electric emission field, which is in fact identical to the drift field within the liquid. It should be remarked that in the cathode gap the drift fields within the liquid $E_{LAr}$ and in the gas phase $E_D$ (see Fig. 1) are different:

$$E_{LAr} = V_C / (d_L + d_G \varepsilon_L / \varepsilon_G)$$
$$E_D = V_C / (d_L \varepsilon_G / \varepsilon_L + d_G) \quad (1)$$

Here $V_C$ is the voltage applied to the cathode gap, $d_L$ and $d_G$ the liquid and gas layer thickness, $\varepsilon_L$ and $\varepsilon_G$ the liquid and gas dielectric constants.

The gain of the two-phase avalanche detector were measured with pulsed X-rays with an amplifier's shaping time of 10 μs, similar to that in our previous works [16]: the gain is defined as the pulse-height of the avalanche (anode) signal from the GEM multiplier divided by that of the calibration signal. The latter was recorded at the first electrode of the first GEM, the cathode gap being operated in an ionization collection mode.

The cryogenic chamber was filled with Ar or Ar+$N_2$, the appropriate gas taken directly from three 40 l bottles permanently connected to the chamber. The gas was purified by flowing through an Oxisorb filter during cooling and heating procedures. The electron life-time in the liquid was larger than 25 μs.

In case of Ar+$N_2$, the $N_2$ concentration in the liquid and gas phase was calculated using Raoult's law [33]:

$$P(N_2 \, in \, gas) = P(N_2 \, saturated) \cdot X(N_2 \, in \, liquid) \quad (2)$$

Here $P(N_2 \, in \, gas)$ is the $N_2$ partial pressure in the gas phase; $P(N_2 \, saturated)$ is the saturated $N_2$ vapour pressure at a given temperature which is equal to 2.1 atm at 84 K [34]; $X(N_2 \, in \, liquid)$ is the $N_2$ concentration in the liquid. In particular, for 1.5% $N_2$ concentration in the gas, the $N_2$ concentration in the liquid at 84 K will be

$$X(N_2\ in\ liquid) = 0.7 atm \cdot 1.5\% / 2.1 atm = 0.5\% \tag{3}$$

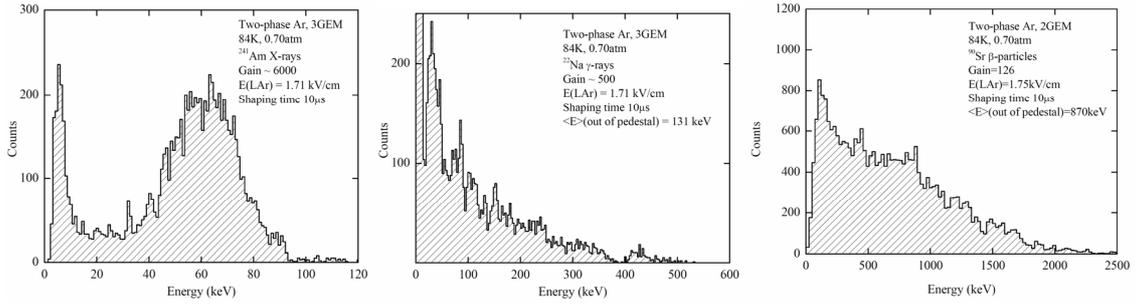

Fig. 2. Energy spectra in a two-phase Ar avalanche detector induced by 60 keV X-rays from $^{241}$Am source (left), 551 keV gamma-rays from $^{22}$Na source (middle) and beta-particles from $^{90}$Sr source (right), at an electric field within the liquid of 1.71-1.75 kV/cm. The spectra were measured at a gain of 6000, 500 and 126 and in 3GEM, 3GEM and 2GEM mode, respectively. The energy scales were calibrated using a 60 keV X-ray line of $^{241}$Am source.

Here we took into account that the saturated vapour pressure in two-phase Ar at 84 K is equal to 0.70 atm [32]. To provide such concentration the required amount of $N_2$ was calculated and doped into the gas system before cooling. Thus at 84 K the two-phase Ar+$N_2$ system was actually Ar+0.5%$N_2$ in the liquid phase and Ar+1.5%$N_2$ in the gas phase.

The detector was irradiated from outside by different ionization sources through two stainless steel windows at the chamber bottom, 100 μm thick and 1 cm diameter each. The energy spectra of ionization sources, as recorded by the two-phase Ar avalanche detector, are shown in Fig.2; the energy scales were calibrated using a linear extrapolation from a 60 keV line of $^{241}$Am source. The ionization sources used were the following: a pulsed X-ray tube providing X-rays with a deposited energy in the range of 30-40 keV, $^{241}$Am source providing a 60 keV X-ray line, $^{22}$Na source providing 511 keV gamma-rays with the average deposited energy of 130 keV and $^{90}$Sr source providing beta-particles with the average deposited energy of 870 keV.

In the latter two cases the average energy was calculated using only the high energy component of the spectrum; the "pedestal" events reflecting the low energy component were disregarded in order to be less sensitive to the detection threshold. To guarantee the average energy value reproducibility, the $^{90}$Sr source position was the same in all the measurements, since otherwise this value would be dependent on the beta-particle range before the cathode. In the case of $^{241}$Am source, only the signals induced by 60 keV X-rays were selected by a trigger in the pulse-shape analysis.

One can see that the average energy deposited in the liquid and recorded by the detector varied by more than an order of magnitude for different ionization sources. We will see in the following, that the average energy deposition will affect the ionization yield from a track in the liquid.

It should be remarked that in the present work the experimental data were obtained in several measurement runs conducted over the course of the year. During this period three different GEM assemblies with CsI photocathode were employed in the cryogenic chamber, with a good reproducibility of results.

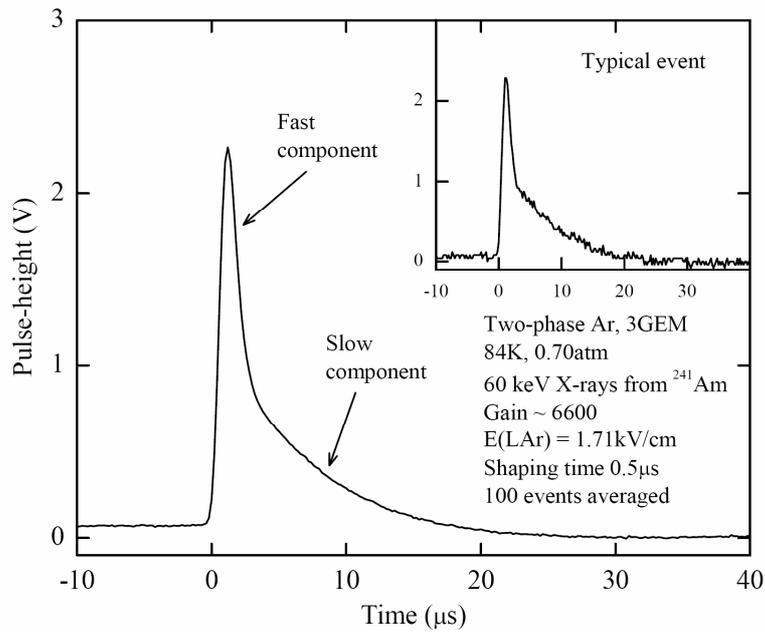

Fig. 3. Anode signal averaged and a typical signal (in the inset) in a two-phase Ar avalanche detector at an electric field within the liquid of 1.71 kV/cm, in 3GEM mode at a gain of 6600, induced by 60 keV X-rays from $^{241}$Am source. The amplifier shaping time is 0.5 μs. Fast and slow components are distinctly seen.

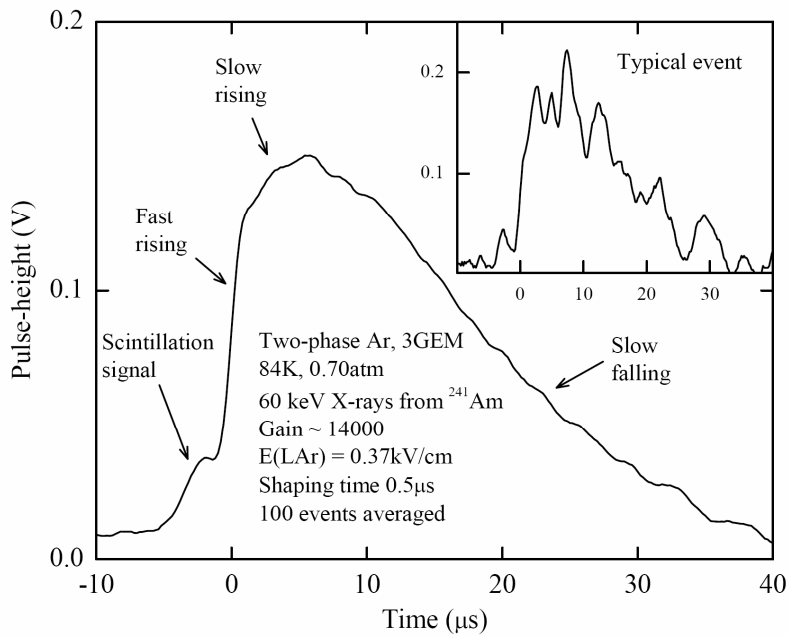

Fig. 4. Anode signal averaged and a typical signal (in the inset) in a two-phase Ar avalanche detector at an electric field within the liquid of 0.37 kV/cm, in 3GEM mode at a gain of 14000, induced by 60 keV X-rays from $^{241}$Am source. The amplifier shaping time is 0.5 μs. The small signal prior to the higher ionization signal is due to primary scintillations in the liquid.

# 3. Electron emission properties of two-phase Ar avalanche detectors

## 3.1 Fast and slow electron emission components

We managed to observe both fast and slow electron emission components in two-phase Ar using a pulse-shape analysis. This is illustrated in Figs. 3 and 4 showing averaged and typical anode signals in a two-phase Ar avalanche detector operated in 3GEM mode, at a higher and lower electric field within the liquid, at 1.71 and 0.37 kV/cm respectively.

In Fig. 3 the fast and slow components are distinctly seen, corresponding to the fast peak at the pulse-rise and to the slow pulse-drop respectively. Since the emission of hot electrons responsible for the fast component takes less than a nanosecond [24], the pulse-shape of the fast component is defined by the amplifier shaping time (0.5 μs), i.e. it should be identical to that of the amplifier calibration signal. The latter was determined by supplying a delta-function-like signal to the preamplifier input, produced by a pulse generator connected through a RC circuit having a small time constant. An example of such a calibration signal identified as a fast component is shown in Fig. 6 (right). On the other hand, the slow component decay time varies between several and several tens microseconds, allowing its time structure to be analyzed in detail.

In Fig. 4 the separation between the fast and slow components is less obvious if at all possible. While the slow component is well defined by the slow falling edge of the anode signal, it is impossible to select the fast component unambiguously: on equal terms it could be either fully suppressed or attributed to the fast rising edge of the signal (see section 3.3 for further details). Anyway one may conclude that the fast and slow components fractions and the slow component decay time strongly depend on the electric field, confirming the statement that these components are governed by the electron emission processes at the liquid-gas interface.

At lower electric fields (Figs. 4) in addition to the ionization signal and prior to it, a fast signal due to primary scintillations in liquid Ar was observed, recorded with CsI photocathode on the first GEM: see Refs. [14] and [15] for more details. The amplitude of this signal, of about 2 photoelectrons at the CsI photocathode for the 60 keV deposited energy [15], was relatively small and thus did not affect the result of the fast and slow components analysis.

It should be remarked that at an amplifier shaping time of 0.5 μs the measurement channel had a limited sensitivity to slow components with time constants larger than 50 μs, if any, due to noises at the signal tail and limited time interval measured. For example, simulations showed that at a drift field of 1.7 kV/cm the secondary slow component with a decay time constant of 50 μs would be observable only if its contribution exceeds 20%.

## 3.2 Physical processes at the liquid-gas interface and thermionic emission models

To understand the fast and slow component origin, the basic physical processes at the liquid-gas interface should be considered. Their simplified model is schematically depicted in Fig. 5; it includes the following steps:

(1) Direct emission of "hot" electrons, heated by an electric field during drifting towards the interface, satisfying the escape cone condition [24],[27]

$$p_Z > (2mV_0)^{1/2}, \tag{4}$$

where $m$ is the electron mass, $p_Z$ is the electron momentum projection to the axis orthogonal to the liquid-gas surface, $V_0$ is the potential barrier height equal to the absolute value of the energy of the ground state of free electrons in the liquid. In liquid Ar this energy is negative (see Fig.

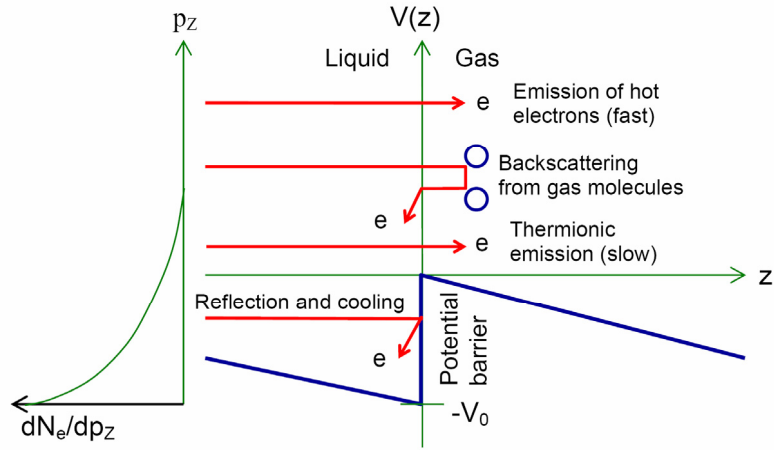

Fig. 5. Potential energy diagram and physical processes at the liquid-gas interface in a two-phase system.

5); its value is not well determined [24],[27]: $-V_0 = -0.065 \div -0.21$ eV. This is a fast process, giving rise to the fast electron emission component.

(2) Reflections of "hot" electrons from the potential barrier not satisfying the escape cone condition, their returns to the barrier due to elastic collisions with the molecules in the liquid and repeated reflections, until the escape cone condition is satisfied and the electrons come out from the liquid [24]. This is a fast process contributing to the fast electron emission component, taking $10^{-9} \div 10^{-10}$ s [24].

(3) Cooling (thermal relaxation) of reflected electrons [27]. This is a fast process, taking $10^{-9} \div 10^{-10}$ s [24].

(4) Thermionic (thermoelectric) emission of "cold" electrons, having overcome a potential barrier [25],[27]. This is a slow process, giving rise to the slow electron emission component, since only energetic electrons from the tail of the Maxwell distribution, having $p_Z > (2mV_0)^{1/2}$, can be emitted.

(5) Backscattering of emitted electrons from the gas molecules to the liquid, followed by electron cooling (step 3). This is a fast process.

It should be remarked that the electron backscattering effect in two-phase systems (step 5) was not taken into account in previous works. On the other hand, the electron backscattering effect for electrons emitted from photocathodes in gas media is well known: see for example a recent review [35]. This effect is particularly strong in noble gases and thus should certainly manifest itself in two-phase noble gas systems. In section 6 it will help us to explain the difference in emission properties of two-phase Ar and Ar+$N_2$.

Let us consider in more detail the thermionic emission process (step 4). In general the emission frequency $\nu$ is proportional to the frequency, with which a reflected electron returns to the barrier [25]:

$$\nu \sim \frac{1}{return\ period} = \frac{1}{\lambda_1 / v_{DL}} \ . \qquad (5)$$

Here $\lambda_1$ is the electron mean free path in the liquid for momentum transfer and $v_{DL}$ is the electron drift velocity in the liquid. Accounting for thermionic emission probability described by a factor $exp(-V_0/k_B T)$, we have [25]:

$$\nu \sim \frac{v_{DL}}{\lambda_1} \exp\left(-\frac{V_0}{k_B T}\right).  \tag{6}$$

Taking into account that the drift velocity in the liquid $v_{DL} = \mu E$, where $\mu$ is the electron mobility and $E$ is the electric field within the liquid, the decay time constant of the slow electron emission component in the frame of *thermionic emission model* will be

$$\tau = 1/\nu \sim \frac{\lambda_1}{\mu E} \exp\left(\frac{V_0}{k_B T}\right).  \tag{7}$$

Thus the slow component decay time can be inversely proportional to the electric field [27]:

$$\tau \sim \lambda_1 / v_{DL} = \lambda_1 / \mu E \Rightarrow \tau = a/E,  \tag{8}$$

where $a$ is some parameter independent of the field. Note that this is valid only if the electron mobility $\mu$=const, which is not always the case.

In Ref. [25] *thermionic emission model with Schottki effect*, i.e. with a decrease of the potential barrier by an electric field, was considered resulting in the following modification of expression (7):

$$\tau \sim \frac{\lambda_1}{\mu E} \exp\left(\frac{V_0 - 2eA_L^{1/2}[1+(A_G/A_L)^{1/2}]E^{1/2}}{k_B T}\right).  \tag{9}$$

$$A_G = \frac{e^2}{16\pi\varepsilon_0\varepsilon_G}\frac{\varepsilon_L - \varepsilon_G}{\varepsilon_L + \varepsilon_G}; \quad A_L = A_G \varepsilon_G / \varepsilon_L$$

Here $\varepsilon_L$, $\varepsilon_G$ и $\varepsilon_0$ are the dielectric constants of the liquid, vapour and vacuum respectively.

Furthermore, the finite electron life-time in the liquid due to attachment by electronegative impurities, *t1*, should be taken into consideration. Then the slow component decay time constant measured in experiment, *t0*, in the frame of *thermionic emission model with electron attachment* will be [27]:

$$t0 = \frac{\tau \cdot t1}{\tau + t1} = \frac{t1}{1+(t1/a)E}$$

$$Y_E = \frac{t1}{\tau + t1}  \tag{10}$$

Here expression (8) was used and $Y_E$ is the electron emission efficiency, which is lower than 100%.

### 3.3 Characteristics of the fast and slow electron emission components

In the frame of the model described in the previous section, the fast and slow component signals should appear in our measurements simultaneously, with the similar fast rising edges defined by the time resolution of the measurement channel. However at lower electric fields (see Fig. 4) in addition to the expected fast rising and slow falling parts of the anode signal, the slow rising part was observed which is beyond the scope of the model. Accordingly we cannot make an unambiguous selection of the fast and slow emission components, since the true shape of the rising edge of the slow component is actually unknown. One can nevertheless assign the limits on the fast and slow components fractions using two extreme approaches.

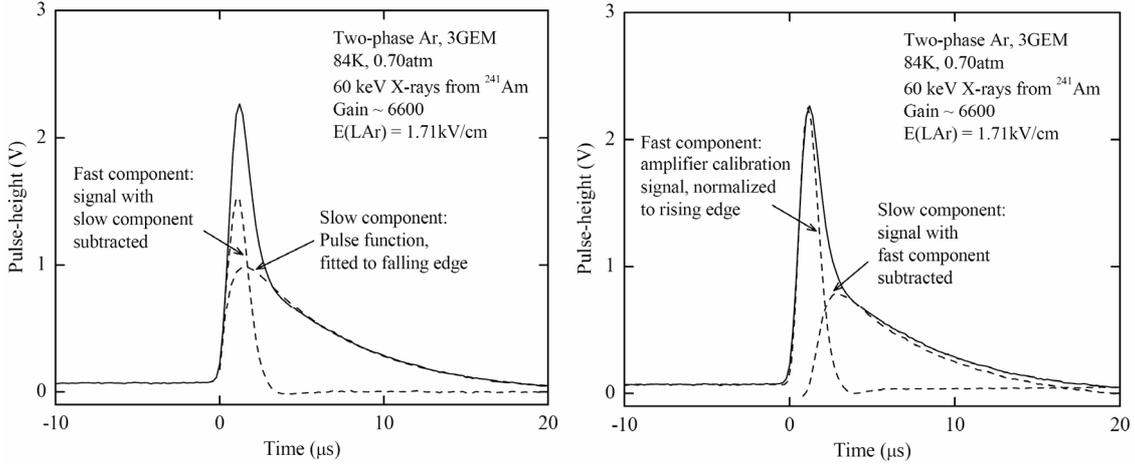

Fig. 6. Fast and slow components selection from the anode signal in a two-phase Ar avalanche detector at an electric field within the liquid of 1.71 kV/cm using two extreme approaches: selecting the slow component by fitting the falling edge of the anode signal with the Pulse function (left); selecting the fast component by normalizing the amplifier calibration signal to the rising edge of the anode signal (right). In both cases the complementary component is obtained by subtracting the selected component from the anode signal. The amplifier shaping time is 0.5 μs.

In the first approach the slow component is selected by fitting the falling edge of the anode signal with a so-called "Pulse" function having two exponential functions, with time constants $t2$ and $t0$ describing the pulse-rise and the pulse-decay respectively (see Fig. 6, left):

$$y = y0 + A\left[1 - \exp\left(-\frac{x-x0}{t2}\right)\right] \exp\left(-\frac{x-x0}{t0}\right) \quad (11)$$

In particular in Figs. 6 (left), the pulse-rise of the slow component with time constant $t2$ was taken to be fast, namely the same as that of the fast component described by the calibration signal, while the decay time $t0$ was a fitting parameter. The fast component is obtained by subtracting the slow component from the anode signal. At lower fields the fast component in this approach completely disappeared, the anode signal being well reproduced exclusively by the slow component.

In the second approach the fast component is selected by normalizing the amplifier calibration signal to the rising edge of the anode signal (see Fig. 6, right): this mimics the fast component contribution. The slow component is obtained now by subtracting the fast component from the anode signal. Here the decay time of the slow component was obtained by fitting it with the Pulse function (11).

The fraction of the fast component is defined as the ratio of areas of the fast component signal to that of the total anode signal ("fast+slow" component signal). In Fig. 7 the limits of this quantity obtained using the two approaches are plotted as a function of the electric field within the liquid. One can see that at a temperature of 84 K the slow component of the electron emission dominated even at higher electric fields (approaching 2 kV/cm), the fast component fraction varying from practically zero to about 30%. This observation contrasts with that of Ref. [24] where the fast component dominated at all fields, the slow component disappearing at fields exceeding 1.5 kV/cm (note that those data were obtained at higher temperature, namely at 90 K, and that the fast component there was not correctly defined due to insufficient time resolution).

The decay time constant of the slow emission component, obtained by fitting with the Pulse function (11), is the same in both approaches. In Fig. 8 it is shown as a function of the electric field within the liquid. One can see that the decay time constant of the slow component varied from 5 to 25 μs in the field range of 0.25-1.71 kV/cm, being larger for lower fields.

The field dependence of the slow component decay time constant shown in Fig. 8 is better compatible with thermionic emission model without electron attachment, described by $t0 \sim 1/E$ dependence. In this case the electron emission efficiency is equal to 100%. At the same time, the data of Fig. 8 can also be fitted by thermionic emission model with electron attachment, described by expression (10), with the electron life-time in the liquid equal to

$$t1 = 45 \pm 20\, \mu s\,. \tag{12}$$

Here the electron emission efficiency is below 100%. It should be remarked that in the present work the measurement accuracy was not enough to distinguish between thermionic emission models without and with Schottky effect, i.e. between expressions (7) and (9).

Our data on slow component decay time constant cannot be compared directly to those of Ref. [25], since the latter were measured at lower electric fields. We can however extrapolate the data of Ref. [25] to our field region using expression (9). This extrapolation predicts the following time constants: 100 μs at 0.37 kV/cm, 40 μs at 0.61 kV/cm and 5 μs at 1.71 kV/cm. These should be compared to the values measured in the present work in Fig. 8 (the data with higher accuracy, obtained with $^{241}$Am source, are only used): 25±2 μs, 13±2 μs and 5.3±1 μs respectively. According to expression (10) the two sets of data might be compatible if the electron life-time in the liquid is of the order of 20-30 μs, which is compatible with that of (12). Consequently, one may conclude that the decay time constants of the present work are also consistent with that of [25] in the frame of thermionic emission model with Schottky effect and electron attachment. It should be noticed that in that case the electron emission efficiency, according to (10), would be lower than 100% at lower electric fields: of about 30% at 0.37 and 0.61 kV/cm.

The general conclusion of this section is that the slow electron emission component in two-phase Ar dominated at fields reaching 2 kV/cm and that its thermionic nature, in terms of the field dependence of the time constant, is confirmed. A particular thermionic emission model cannot be fixed however.

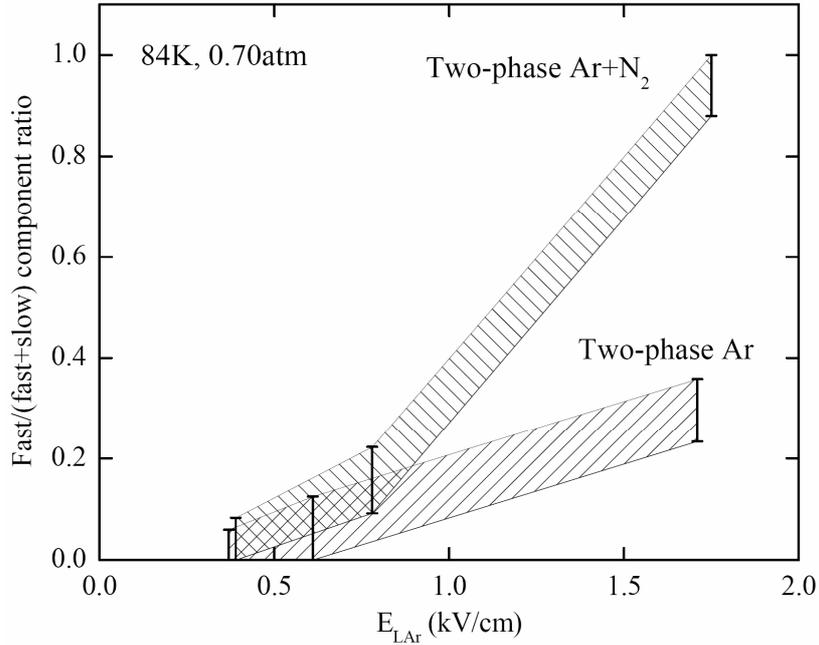

Fig. 7. Fast electron emission component fraction as a function of the electric field within the liquid in two-phase Ar and two-phase Ar+$N_2$ (0.5% in the liquid, 1.5% in the gas) at 84 K and 0.7 atm. Shown are the limits of the fraction value. The signals are induced by 60 keV X-rays from $^{241}$Am source in two-phase Ar and by beta-particles from $^{90}$Sr source in two-phase Ar+$N_2$.

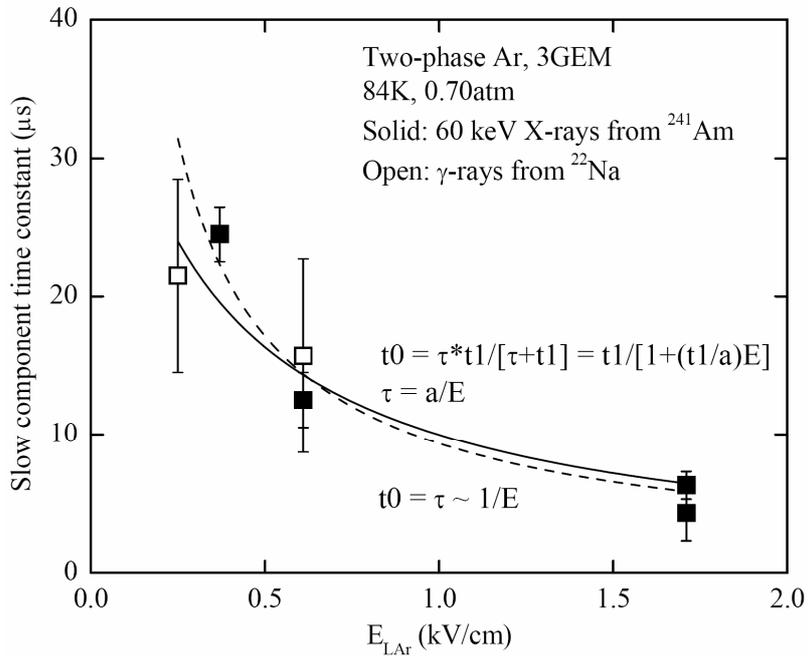

Fig. 8. Decay time constant of the slow electron emission component (*t0*) in two-phase Ar as a function of the electric field within the liquid. The signals are induced by 60 keV X-rays from $^{241}$Am source and gamma-rays from $^{22}$Na source. The data points are fitted by thermionic electron emission model without (dashed line) and with (solid line) electron attachment in the liquid.

## 4. Electron emission properties of two-phase Ar+$N_2$ avalanche detectors

The Ar+$N_2$ two-phase system was prepared as described in section 2, providing $N_2$ concentration of 0.5% in the liquid and 1.5% in the gas phase, at 84 K and 0.70 atm. Fig. 9 illustrates the fact that gain characteristics of the two-phase Ar+$N_2$ avalanche detector did not significantly differ from those of two-phase Ar. It should be remarked that in terms of the maximum avalanche gain the Ar+$N_2$ gas mixture can be even better than that of Ar [36]: at room temperature and atmospheric pressure the triple-GEM gain in this mixture reached $10^5$. Moreover, the avalanche signal in Ar+$N_2$ mixture is rather fast [36]: in Ar+1.3%$N_2$ the anode pulse width from the triple-GEM was only 26 ns (FWHM), which is comparable to that of Ar+10%$CH_4$.

Figs. 10 and 11 demonstrate the most remarkable observation of the present work: a disappearance of the slow electron emission component in two-phase Ar+$N_2$ at higher electric fields. In Fig. 10 typical anode signals in two-phase Ar and Ar+$N_2$ avalanche detectors are compared in 2GEM mode at a field of 1.75 kV/cm. Fig. 11 shows some more signals in two-phase Ar+$N_2$ obtained in the other measurement run. One can see that at this field value the slow component indeed disappeared in two-phase Ar+$N_2$, the fast component being substantially enhanced. The latter is clear if to compare the fast component pulse-heights in two-phase Ar and Ar+$N_2$, with an account of the gain.

The signals in this section were induced by beta-particles from $^{90}$Sr source; their pulse-height spectra are shown in the insets of Fig. 10. One can see that the spectra in two-phase Ar and Ar+$N_2$ detectors have rather similar shapes. An important conclusion can be derived from this observation: the emitted electrons were not lost when doping Ar with $N_2$; consequently, all the electrons turned into the fast component.

At lower electric fields the slow component in two-phase Ar+$N_2$ reappeared: see Fig. 12. Here the signals look rather similar to those of two-phase Ar; their amplitudes decreased compared to higher fields as seen from pulse-height spectra in the insets.

Using signal treatment procedure described in section 3.3, the contributions of the fast and slow components were quantified: see in Fig. 7. One can see that at fields exceeding 1.5 kV/cm the fast electron emission component dominates in two-phase Ar+$N_2$, its contribution approaching 100%, which is in contrast to two-phase Ar. On the other hand, at fields lower than 1 kV/cm the fast component fraction in two-phase Ar+$N_2$ is small, approaching that of two-phase Ar.

It is interesting that in two-phase Ar+$N_2$ the primary scintillation signal (small signal prior to ionization signal) was not observed (see Fig. 12), in contrast to two-phase Ar. This is obviously due to the fact that the scintillations in liquid Ar+$N_2$ are almost fully suppressed as was reported elsewhere [29],[30]: compared to liquid Ar their intensity decreased by 2-3 orders of magnitude.

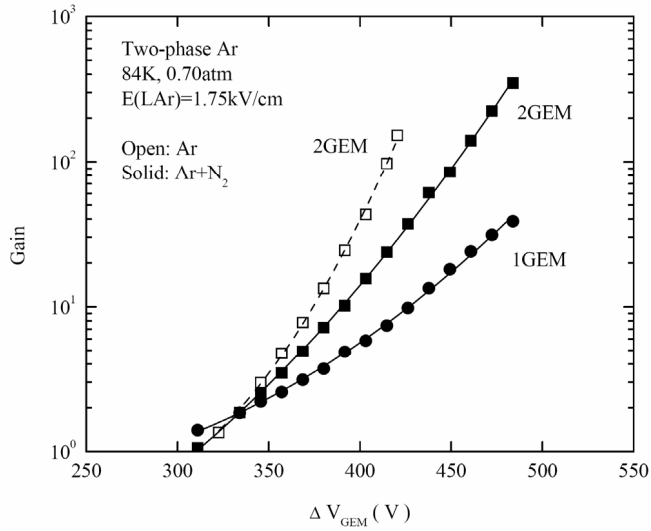

Fig. 9. Gain-voltage characteristics of a two-phase Ar+$N_2$ (0.5% in the liquid, 1.5% in the gas) avalanche detector in 1GEM and 2GEM mode, at an electric field within the liquid of 1.75 kV/cm. For comparison the gain characteristic of a two-phase Ar detector in 2GEM mode is shown. The maximum gains are not reached.

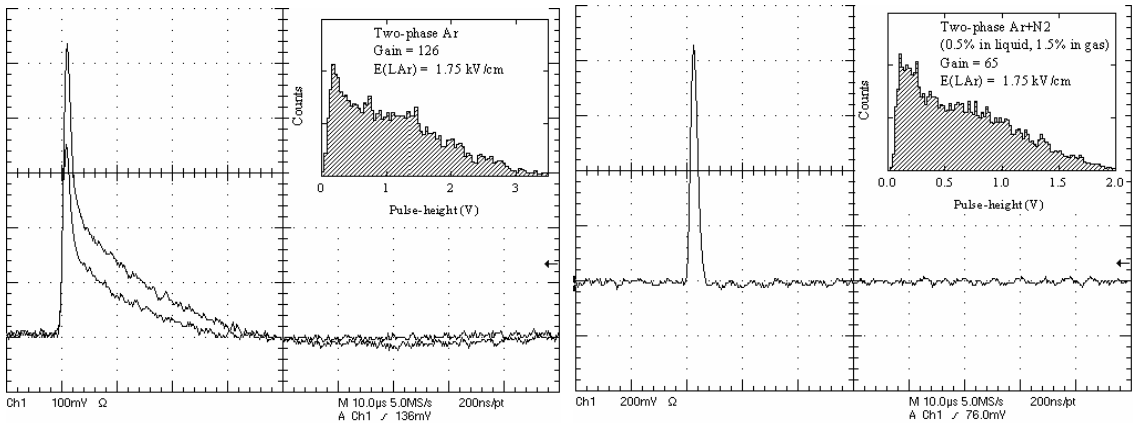

Fig. 10. Typical anode signals in a two-phase Ar avalanche detector (left) and in a two-phase Ar+$N_2$ (0.5% in the liquid, 1.5% in the gas) avalanche detector (right), in 2GEM mode at a gain of 126 and 65 respectively, induced by beta-particles from $^{90}$Sr source at an electric field within the liquid of 1.75 kV/cm. The amplifier shaping time is 0.5 µs. The horizontal scale is 10 µs/div. In the insets, pulse-height spectra are shown at amplifier shaping time of 10 µs.

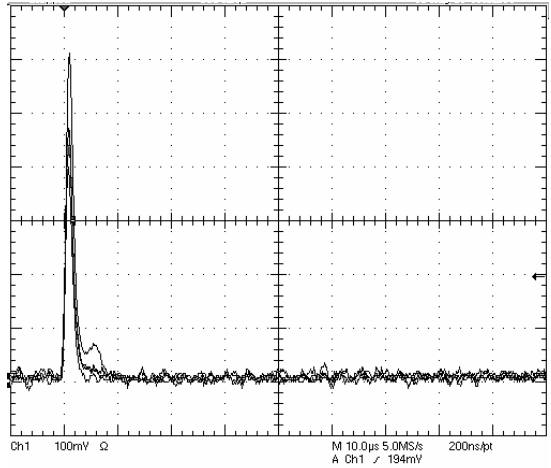

Fig. 11. Typical anode signals in a two-phase Ar+$N_2$ (0.5% in the liquid, 1.5% in the gas) avalanche detector, in 2GEM mode at a gain of 224, induced by beta-particles from $^{90}$Sr source at an electric field within the liquid of 1.75 kV/cm. The amplifier shaping time is 0.5 µs. The horizontal scale is 10 µs/div.

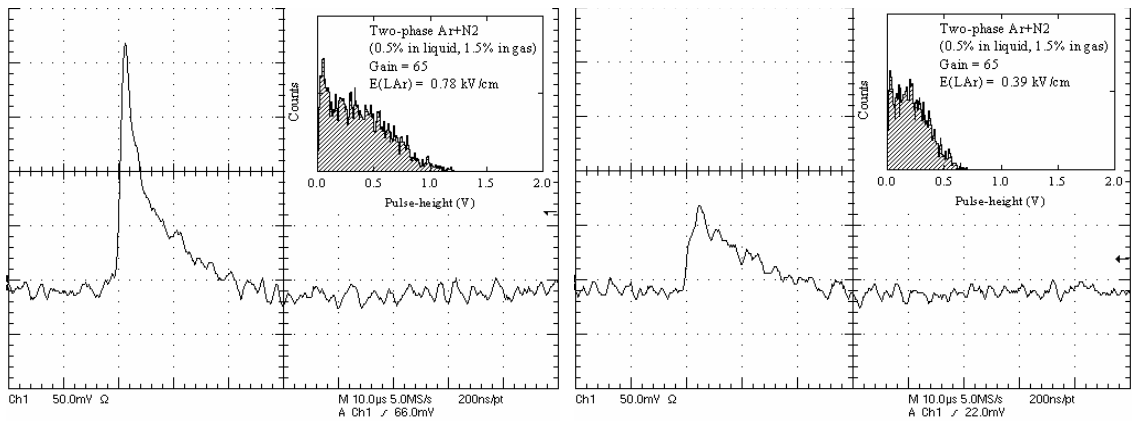

Fig. 12. Typical anode signals in a two-phase Ar+$N_2$ (0.5% in the liquid, 1.5% in the gas) avalanche detector at an electric field within the liquid of 0.78 (left) and 0.39 kV/cm (right), in 2GEM mode at a gain of 65, induced by beta-particles from $^{90}$Sr source. The amplifier shaping time is 0.5 µs. The horizontal scales are 10 µs/div. In the insets, pulse-height spectra are shown at amplifier shaping time of 10 µs.

## 5. Ionization yield and electron emission efficiency in two-phase systems

A dependence of the ionization yield on the electric field in two-phase Ar and Ar+$N_2$ is illustrated in Figs. 13 and 14. Here the ionization yield means the amount of charge extracted from the liquid into the gas phase normalized to the primary ionization charge created in the liquid by a particle. The ionization yield in a two-phase system ($Y$) is combined from the ionization yield from a track in the liquid ($Y_L$) and the electron emission efficiency at the liquid-gas interface ($Y_E$):

$$Y = Y_L \cdot Y_E \quad . \tag{13}$$

Both quantities, $Y_L$ and $Y_E$, depend on the electric field. The ionization yield from a track in addition depends on the ionization density due to electron-ion recombination, i.e. on the deposited energy and the ionizing power of the radiation. In recombination model it is parameterized as follows [28]:

$$Y_L = \frac{1}{1+k/E} \quad . \tag{14}$$

Here parameter $k$ depends on the ionization density and the medium; it is larger for densely ionization.

Fig. 13 shows the field dependence of the quantity proportional to the number of emitted electrons and thus to the ionization yield in two-phase Ar and Ar+$N_2$, when irradiated with beta-particles from $^{90}$Sr source. This quantity is just the average anode amplitude from the GEM multiplier divided by its gain. It was obtained from the pulse-height spectra measured at an amplifier shaping time of 10 μs disregarding the low energy component of the beta-particle spectra as explained in section 2; examples of the spectra are presented in the insets of Figs. 10 and 12.

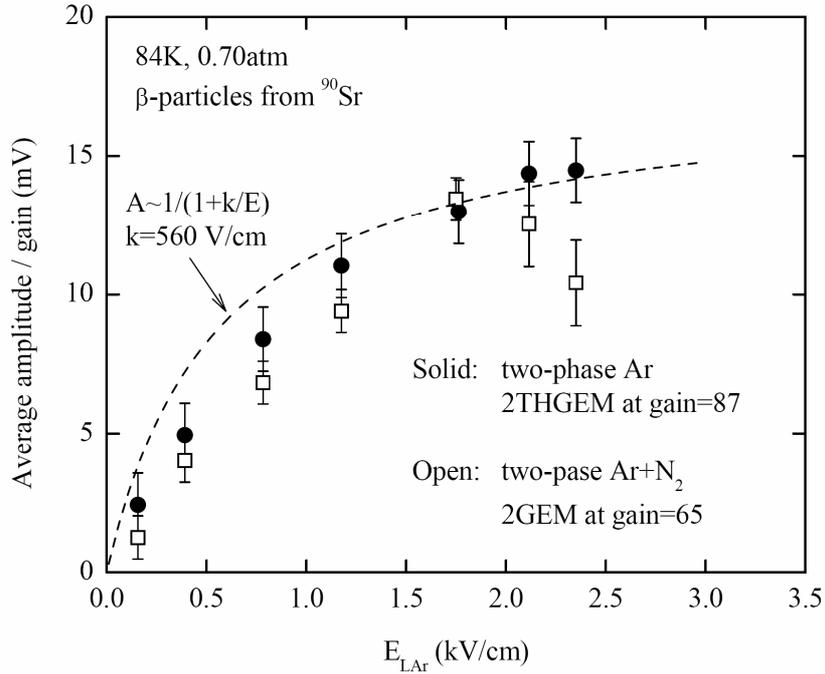

Fig. 13. Field dependence of the relative ionization yield in two-phase Ar and Ar+$N_2$ (0.5% in the liquid, 1.5% in the gas), when irradiated with beta-particles from $^{90}$Sr source. Shown is the average anode spectrum amplitude from the double-GEM divided by gain as a function of the electric field within the liquid. Dashed curve: recombination model function A~1/(1+k/E) at k=560 V/cm normalized to Ar data at E>1.75kV/cm. The amplifier shaping time is 10 μs.

One can see from Fig. 13 that the ionization yield in two-phase Ar and Ar+N$_2$ is about the same. This supports the statement that the emitted electrons are not lost when doping two-phase Ar with N$_2$. Moreover, at fields exceeding 1.5 kV/cm practically all electrons in Ar+N$_2$ are emitted in the fast emission component: see Figs. 7, 10 and 11. Thus doping Ar with N$_2$ may provide the fast and lossless electron emission.

At fields exceeding 2 kV/cm the electron emission efficiency in two-phase Ar was reported to approach 100% [24]. Consequently, if to normalize the recombination model function (14) to Ar data at higher fields in Fig. 13, one could select the contributions of the ionization yield from a track and the electron emission efficiency at lower fields, according to (13). Here parameter $k$ was taken equal to a value $k$=560 V/cm, that was measured elsewhere for the similar ionization source, namely for 1 MeV electrons in liquid Ar [28]. Surprisingly the field dependence of the experimental data at lower fields tends to be well described by the recombination model function, the electron emission efficiency being estimated to exceed 50% at all fields.

Thus we come to a conclusion that the electric field dependence of the ionization yield in two-phase Ar systems is mostly governed by the ionization yield from a track in the liquid described by recombination model, rather than by the electron emission efficiency. This is totally different from two-phase Kr and Xe systems, where the electron emission efficiency has specific threshold behaviour as a function of the field, being substantially lower than 50% even

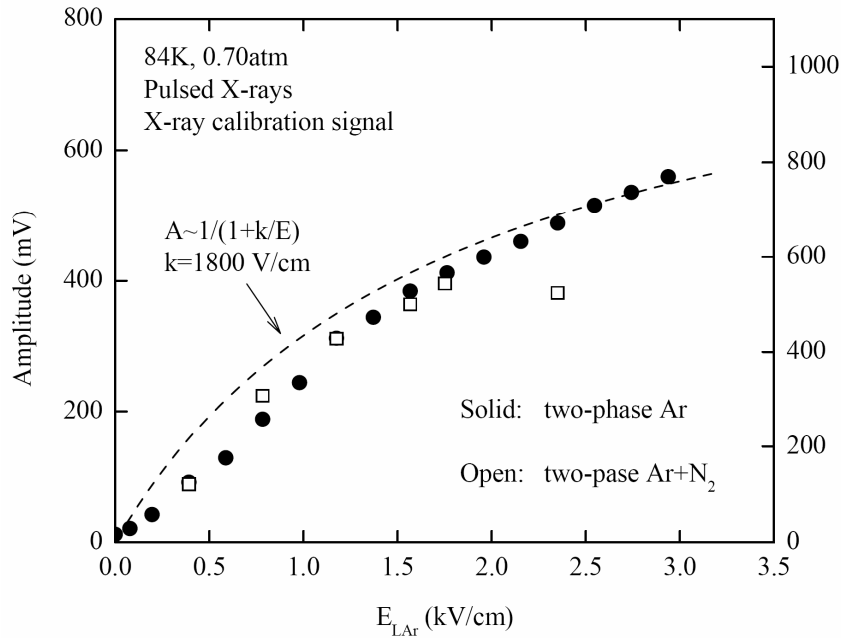

Fig. 14. Field dependence of the relative ionization yield in two-phase Ar and Ar+N$_2$ (0.5% in the liquid, 1.5% in the gas), when irradiated with pulsed X-rays. Shown is the calibration signal amplitude as a function of the electric field within the liquid. Ar data: left scale. Ar+N$_2$ data: right scale. Dashed curve: recombination model function A~1/(1+k/E) at k=1800 V/cm normalized to Ar data at E>2.5kV/cm. The amplifier shaping time is 10 μs.

at higher electric fields [28].

The above assumption is also supported by the field dependence of the relative ionization yield in two-phase Ar and Ar+$N_2$ when irradiated with pulsed X-rays: see Fig. 14. Pulsed X-rays, with the average deposited energy of 30-40 keV, produce more dense ionization in the liquid than that of beta-particles. In accordance with recombination model this should result in more flat field dependence, which is indeed observed in experiment (Fig. 14). Similarly to Fig. 13, the recombination function (14) was normalized to Ar data at higher fields using the parameter value $k$=1800 V/cm, measured elsewhere for 21 keV X-rays in liquid Xe [28]. We have the right here to apply the liquid Xe parameter to liquid Ar data since both liquids have close values of $k$ [28]. One can see that the field dependence again tends to be well described by the recombination model function, the electron emission efficiency exceeding 50% at all fields.

It should be remarked that the calibration signal pulse-heights presented in Fig. 14 were measured in two-phase Ar and Ar+$N_2$ under different conditions. Therefore only the relative field dependences can be compared here. Nevertheless, one may conclude that the field dependence of the ionization yield in two-phase Ar and Ar+$N_2$ when irradiated with X-rays is much about the same, similarly to a conclusion derived from Fig. 13 when irradiated with beta-particles.

At the end of the section let us draw an attention to a possible mysterious behaviour of the ionization yield in two-phase Ar+$N_2$ at fields exceeding 2.3 kV/cm (see Figs. 13 and 14): the ionization yield seems to start decreasing. Though such a decrease is observed only for the last data point, it might hardly be an artefact since it is presented in two different sets of measurements: using beta-particles and X-rays. At the moment we cannot offer a satisfactory explanation of this observation.

## 6. Accounting for electron backscattering effect

As mentioned in section 3.2, the effect of electron backscattering from the gas molecules into the liquid has not been accounted for in two-phase systems before the present study. In this section we further develop the electron emission model described in section 3.2, taking into consideration this effect to explain the difference between two-phase Ar and Ar+$N_2$ systems.

Electron backscattering to the liquid is induced mostly by elastic collisions of electrons with the gas molecules. Then the electron emission can be considered as a two-stage process consisting of electron overcoming a potential barrier at the liquid-gas interface, with the barrier penetration factor $F_{LG}$, and electron transmission through the gas molecules near the interface, with the electron transmission efficiency $F_G$. With regard to photoemission from solid photocathodes in gas media, the latter factor is also often called the photoelectron extraction efficiency.

Similarly to (5) the frequency of such a two-stage process will be proportional to the electron drift velocity at each stage, i.e. to the drift velocity in the liquid $v_{DL}$ and in the gas $v_{DG}$, divided by the electron mean free path for momentum transfer in the liquid $\lambda_{1L}$ and in the gas $\lambda_{1G}$, combined with the appropriate factors:

$$\nu \sim \frac{v_{DL}}{\lambda_{1L}} \cdot F_{LG} \cdot \frac{v_{DG}}{\lambda_{1G}} \cdot F_G \ . \tag{15}$$

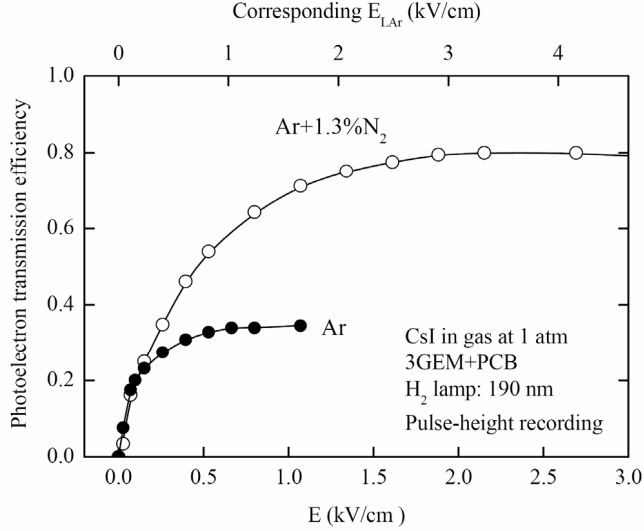

Fig. 15. Efficiency of photoelectron transmission from CsI photocathode into gas as a function of the electric field in Ar and Ar+1.3%$N_2$ at 1 atm and room temperature, at a wavelength of 190 nm. The data were obtained with semitransparent CsI photocathode coupled to a triple-GEM multiplier and irradiated with a pulsed hydrogen lamp, in a pulse-height recording mode. The pulse-height data were normalized to the gas/vacuum current ratio in the appropriate gas taken from [36]. The top scale represents the electric fields within the liquid in two-phase Ar, corresponding to similar $E/N$ values in the gas phase.

This expression can be applied to both fast and slow electron emission processes. In particular it is proportional to the fast emission component fraction measured in experiment.

Factor $F_{LG}$ is defined by the escape cone condition (4) and thus is the same in Ar and Ar+$N_2$ for small $N_2$ concentrations. On the contrary, factor $F_G$ strongly depends on the gas medium and thus can explain the difference between Ar and Ar+$N_2$. In particular at higher fields $F_G$ is larger in Ar+$N_2$ as compared to Ar, resulting in enhancement of the fast emission component and appropriate reduction of the slow component in two-phase Ar+$N_2$.

This assumption is supported by the studies of the photoelectron backscattering effect with CsI photocathodes [36],[37]. Indeed, at higher fields the photoelectron transmission efficiency in gaseous Ar+$N_2$ was measured to be substantially larger than in Ar: see Fig. 17 in Ref. [36] and Fig. 2 in Ref. [37]. See also Fig. 15 of the present work where the data obtained by one of us (A.B.) in the course of the study of CsI photocathodes [35],[36] are presented.

To translate the electric fields values in these figures to those within liquid Ar used in the present work one should multiply them by a factor (0.7atm/$\varepsilon_L$)*(295K/84K), in order to obtain similar values of the reduced electric field $E/N$ in the gas phase of two-phase Ar. In particular, a field value of 1.1 kV/cm in gaseous Ar at 1 atm and room temperature corresponds to a value of 1.75 kV/cm in liquid Ar at 0.7 atm and 84 K. One can see from Fig. 17 of [36] and Fig. 15 of the present work that at such a field value the photoelectron transmission efficiency in Ar+1.3%$N_2$ is above 70%, while in Ar it is below 40%. We assume that this difference is enough to explain the suppression of the slow electron emission component observed in two-phase Ar+$N_2$ at higher fields. On the other hand, at fields lower than 0.5 kV/cm in gaseous Ar, corresponding to the fields within liquid Ar lower than 0.8 kV/cm, the difference between Ar and Ar+N$_2$ in terms of the backscattering effect disappears, which may explain the reappearing of the slow component in two-phase Ar+N$_2$.

If the assumption of the role of the backscattering effect is true, the cold electrons and consequently the slow electron emission component are resulted from electron backscattering from the gas molecules, rather than from electron reflection from the potential barrier at the interface. That means that the slow electron emission component in two-phase systems appears whenever the backscattering effect in the gas phase is strong. And vice versa, to suppress the slow electron emission component and enhance the fast component one should dope noble gases with molecular additives reducing the electron backscattering effect, such as N$_2$, CH$_4$ and CF$_4$, or to operate at higher electric fields where elastic scattering from the gas molecules is taken over by inelastic collisions (see review [35] for more details on this subject).

## 7. Conclusions

Electron emission properties of two-phase avalanche detectors were studied in Ar and Ar+N$_2$ (0.5% in the liquid and 1.5% in the gas phase). The detectors investigated comprised a liquid layer followed by a multi-GEM multiplier operated in the saturated vapour above the liquid phase, at 84 K and 0.70 atm.

Two components of the electron emission through the liquid-gas interface were observed directly: fast and slow. In Ar, the slow emission component dominated even at higher fields, reaching 2 kV/cm. In Ar+N$_2$ on the contrary, the fast emission component dominated at higher fields, exceeding 1.5 kV/cm, the slow component being disappeared. Such behaviour is explained by the electron backscattering effect in the gas phase: the slow electron emission component in two-phase systems is supposed to appear whenever the backscattering effect is strong.

The slow component decay time constant was inversely proportional to the electric field, which is compatible with thermionic emission model. The electron emission efficiencies in two-phase Ar and Ar+N$_2$ were estimated to be close to each other, their absolute values exceeding 50% even at lower electric fields.

Our general conclusion is that two-phase Ar+N$_2$ avalanche detectors might be superior to those of other noble gases. They may have fast signals due to the absence of the slow electron emission component and fast avalanche signals, relatively low electric fields needed for efficient electron emission from the liquid and high avalanche gain in the gas phase. One should have in view however that the scintillations in liquid Ar+N$_2$ are suppressed. Such detectors might be relevant to those experiments where only the ionization signal is recorded, in particular to coherent neutrino-nucleus scattering experiments and large-scale neutrino detectors.

## 8. Acknowledgements

We are grateful to R. Snopkov and A. Chegodaev for their help in the development of the experimental setup.

# References

[1] B.A. Dolgoshein et al., *New method of registration of ionizing-particle tracks in condensed matter*, JETP Lett. 11 (1970) 513.

[2] B.A. Dolgoshein et al., *Electronic particle detection method for two-phase systems: liquid – gas* (in Russian), Phys. Elem. Part. Atom. Nucl. 4 (1973) 167.

[3] E. Aprile et al., *The XENON dark matter search: status of XENON10*, J. Phys. Conf. Ser. 39 (2006) 107.

[4] P. Benetti et al., *First results from a dark matter search with liquid argon at 87-K in the Gran Sasso underground laboratory*, Astropart. Phys. **28** (2008) 495 [astro-ph/0701286].

[5] D. Akimov, Detectors for Dark Matter search (review), Nucl. Instrum. Meth. A 598 (2009) 275, and references therein.

[6] C. Hagmann and A. Bernstein, *Two-phase emission detector for measuring coherent neutrino-nucleus scattering*, IEEE Trans. Nucl. Sci. 51 (2004) 2151.

[7] C. D. Winant et al., *Dual-Phase Argon Ionization Detector for Measurement of Coherent Elastic Neutrino Scattering and Medium-Energy Nuclear Recoils*, IEEE Nucl. Sci. Symp. Conf. Rec., N42-4, p. 2110, (2007).

[8] A. Buzulutskov et al., *First results from cryogenic avalanche detectors based on gas electron multipliers*, IEEE Trans. Nucl. Sci. NS-50 (2003) 2491.

[9] A. Buzulutskov, *Radiation detectors based on gas electron multipliers (Review)*, Instr. Exp. Tech. 50 (2007) 287, and references therein.

[10] F. Sauli, *GEM: A new concept for electron amplification in gas detectors*, Nucl. Instrum. Meth. A 386 (1997) 531.

[11] C. Shalem et al., *Advances in thick GEM-like gaseous electron multipliers — Part I: atmospheric pressure operation*, Nucl. Instrum. Meth. A 558 (2006) 475.

[12] V. Peskov et al., *Development and First Tests of GEM-Like Detectors With Resistive Electrodes*, IEEE Trans. Nucl. Sci. NS-54 (2007) 1784.

[13] A. Bondar et al., *Two-phase argon and xenon avalanche detectors based on Gas Electron Multipliers*, Nucl. Instrum. Meth. A 556 (2006) 273.

[14] A. Bondar et al., *First results of the two-phase argon avalanche detector performance with CsI photocathode*, Nucl. Instrum. Meth. A 581 (2007) 241.

[15] A. Bondar et al., *Recent results on the properties of two-phase argon avalanche detectors*, Nucl. Instrum. Meth. A 598 (2009) 121.

[16] A. Bondar et al., *Thick GEM versus thin GEM in two-phase argon avalanche detectors*, JINST 3 (2008) P07001.

[17] P.K. Lightfoot et al., *Optical readout tracking detector concept using secondary scintillation from liquid argon generated by a thick gas electron multiplier*, JINST 4 (2009) P04002.

[18] A. Badertscher et al., *Construction and operation of a Double Phase LAr Large Electron Multiplier Time Projection Chamber*, E-print arXiv:0811.3384v1, 2008.

[19] D. Akimov et al., *Detection of reactor antineutrino coherent scattering off nuclei with a two-phase noble gas detector*, JINST 4 (2009) P06010.

[20] A. Rubbia, *ArDM: a ton-scale liquid Argon experiment for direct detection of Dark Matter in the Universe*, J. Phys. Conf. Ser. 39 (2006) 129.

[21] D. Angeli at al., *Towards a new Liquid Argon Imaging Chamber for the MODULAr project*, JINST 4 (2009) P02003.

[22] F. Balau et al., *GEM operation in double-phase xenon*, Nucl. Instrum. Meth. A 598 (2009) 126.

[23] Y.L. Ju et al., *Cryogenic design and operation of liquid helium in an electron bubble chamber towards low energy solar neutrino detectors*, Cryogenics 47 (2007) 81.

[24] E.M. Gushchin et al., *Hot electron emission from liquid and solid argon and xenon*, Sov. Phys. JETP 55 (1982) 860.

[25] A.F. Borghesani et al., *Electron transmission through the Ar liquid-vapor interface*, Phys. Lett. A 149 (1990) 481.

[26] R. Galea et al., *Charge transmission through liquid neon and helium surfaces*, JINST 2 (2007) P04007, and references therein.

[27] A.I. Bolozdynya, *Two-phase emission detectors and their applications*, Nucl. Instrum. Meth. A 422 (1999) 314.

[28] A.S. Barabash, A.I. Bolozdynya, *Liquid ionization detectors*, Energoatomizdat, Moscow, 1993 (in Russian).

[29] S. Himi et al., *Liquid and solid argon and nitrogen-doped liquid and solid argon scintillatores*, Nucl. Instrum. Meth. 203 (1982) 153.

[30] R. Acciarri et al., *Effects of Nitrogen contamination in liquid Argon*, E-print arXiv:0804.1217, 2008.

[31] A. Ereditato et al., *Study of ionization signals in a TPC filled with a mixture of liquid Argon and Nitrogen*, JINST 3 (2008) P10002.

[32] V.G. Fastovski, A.E. Rovinski, Y.V. Petrovski, *Inert gases*, Atomizdat, Moscow, 1964 (in Russian).

[33] W.J. Moore, Physical Chemistry, Prentice-Hall, NJ, 1972.

[34] *Physical quantities. Reference book*, Eds. I.S. Grigorieva and E.Z. Meilikhova, Ehergoatomizdat, Moscow, 1991 (in Russian).

[35] A. Buzulutskov, *Gasesous photodetectors with solid photocathodes*, Phys. of Part. and Nucl. 39 (2008) 424.

[36] A. Buzulutskov et al., *The GEM photomultiplier operated with noble gas mixtures*, Nucl. Instrum. Meth. A 443 (2000) 164.

[37] A. Breskin et al., *GEM photomultiplier operation in $CF_4$*, Nucl. Instrum. Meth. A 483 (2002) 670.